\documentstyle[prd,aps,epsf,epsfig]{revtex} %c2
\begin{document}
\draft
\twocolumn[\hsize\textwidth\columnwidth\hsize\csname@twocolumnfalse\endcsname%c2
%:::::::::::::::::::::::::::::::::::::::::::::::::::::::::::::::::::::::
%
\title{Spiral cracks in drying precipitates}
%:::::::::::::::::::::::::::::::::::::::::::::::::::::::::::::::::::::::
\vspace{0.5cm}
\author{Z. N\'eda$^a$, K.-t. Leung$^b$, L. J\'ozsa$^c$ and M. Ravasz$^a$}
\address{$^a$ Babe\c{s}-Bolyai University, Dept. of Physics, RO-3400, Cluj, Romania\\ 
$^b$ Institute of Physics, Academia Sinica, Taipei, Taiwan 11529, R.O.C.\\
$^c$ Babe\c{s}-Bolyai University, Dept. of Chemistry, RO-3400, Cluj, Romania}

\maketitle
\centerline{\small (Last revised \today)}

\begin{abstract}
We investigate the formation of spiral crack patterns 
during the desiccation of thin layers of precipitates in contact with 
a substrate. This symmetry-breaking fracturing mode is found to
arise naturally not from torsion forces, but from a propagating stress 
front induced by the fold-up of the fragments. 
We model their formation mechanism using a coarse-grain
model for fragmentation and successfully reproduce the spiral cracks.
Fittings of experimental and simulation data show that the spirals
are logarithmic, corresponding to constant deviation from a 
circular crack path.  
Theoretical aspects of the logarithmic spirals are discussed.  
In particular we show that this occurs generally when the 
crack speed is proportional to the propagating speed of stress front.
\end{abstract}

\vspace{2pc}
]%c2

\vspace{1cm}

%:::::::::::::::::::::::::::::::::::::::::::::::::::::::::::::::::::::::
% Text begins:
%:::::::::::::::::::::::::::::::::::::::::::::::::::::::::::::::::::::::

Fracture of solids produces a large variety of fascinating patterns. 
Straight and wiggling cracks in fragmented dried out fields, 
rocks, tectonic plates and paintings, and the self-affine 
fractured surfaces are just a few well studied examples.
Understanding fracture and fragmentation phenomena is of increasing interest
in physics and engineering \cite{fra_gen}. Many recent studies have analyzed the
morphology of fractured surfaces and fracture lines \cite{recent}, most of which
showing a cellular and hierarchical pattern. Concomitantly, successful
models have been proposed to describe crack propagation and to reproduce
the observed structures \cite{models}.

Apart from the cellular type,
spiral, helical and in general smoothly curving fracturing modes are
known in material science \cite{curving}. Most of the known spiral cracks are
either due to imposed torsion (twist) as in the spiral fracture of the
tibia \cite{tibia}, or due to geometric constraints as in the
fracture of pipes \cite{pipes}. Spiral cracks can, however, also arise in 
situations where no obvious twisting is applied, so that the symmetry is
spontaneously broken.
An example of this  
was recently given by Hull \cite{Hull} in the study of 
the shrinkage of silica based sol-gel.
Similar structures were reported by us for the post-fragmentation 
process of a thin layer of drying precipitate \cite{nature}.
The formation of spiral cracks under specific 
conditions was also recently considered by Xia and Hutchinson \cite{xia}. 
In this letter,
the mechanism leading to this special cracking mode shall be 
investigated and modeled. We shall discuss some mathematical properties in
the shape of the spiral and  reproduce them by computer simulations.

The experimental conditions leading to such structures are simple,
one can do that without a laboratory. 
The first step is to produce by chemical reactions a
fine suspension of precipitate. The spiral cracks are
not restricted to one peculiar material, as we obtain them
with different compounds, including nickel phosphate 
($Ni_3(PO_4)_2$, from the reaction between $NiSO_4$ and $Na_3PO_4$),
ferric ferrocyanide ($Fe_4[Fe(CN)_6]_3$, from 
$K_4[Fe(CN)_6]$ and $FeCl_3$) and ferric hydroxide ($Fe(OH)_3$ from
$FeCl_3$ and $NaOH$). The reacting salts are 
diluted in distiled water to concentrations between $0.3-10\%$ for
all reactions. 
Mixing the two reacting solutions produces the desired compound.
The solution is then left to sediment and
the dissolved ions ($Na^+$, $K^+$, $Cl^-$, $SO_4^-$,...) are removed
by rinsing with distiled water. 
This gives us an aqueous suspension, which is
then poured into a Petri-dish and let dry. 
During drying a thin solidified layer is formed 
which is then fragmented into isolated parts (Fig.~1a).
\begin{figure}[h]
\epsfig{figure=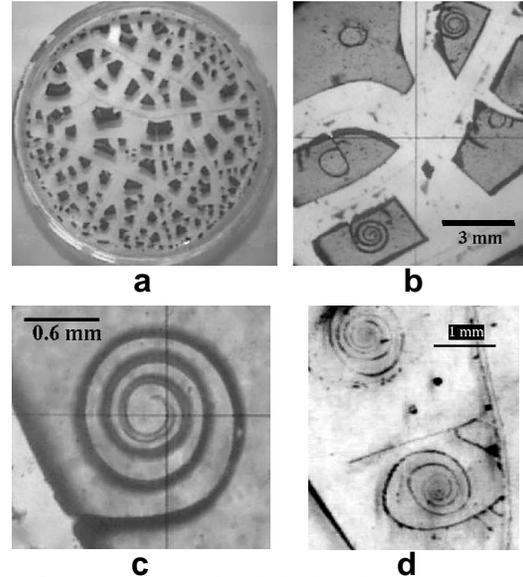,width=2.7in,angle=-0}
\caption{Fracture of nickel phosphate precipitate at different length scales.
(a) Typical fragmentation pattern inside a Petri-dish, 
(b) spiral and circular structures inside the fragments, 
(c) close-up of a spiral crack and (d) traces of the spiral cracks
left on the glass surface.
}
\label{fig1}
\end{figure}
For fine and thin precipitate  
(grain-size smaller than a few hundreds of nm and layer thickness between 0.2 to 0.5 mm), 
regular spirals as well as other smoothly curving cracks 
(such as circles, ellipses, and intersecting arcs)
finally show up {\em inside\/}  the fragments (Fig.~1b). 
Depending on the grain size
and layer thickness the size of these fascinating structures varies 
widely, ranging from several hundred microns to about 5mm. 
Easily overlooked by naked eye, 
they are revealed under a microscope (Fig.~1c). 
Moreover, after removal of the layer, they leave beautiful marks 
on the glass surface (Fig.~1d). 

By digitally analyzing their shape we found that all those spirals 
are approximately logarithmic, i.e., 
described by the functional form
\begin{equation}
r(\theta)=r_0 e^{k\theta},
\label{eq:r_theta}
\end{equation}
where $r_0$ and $k$ are constants, 
with $k$ determining the overall tightness of the spiral.
$r$ and $\theta$ are the polar coordinates in plane
(in our convention $\theta$ can take any real value, not restricted
to $[0,2\pi]$.) 
A characteristic fit is presented in Fig.~2, showing
their logarithmic nature. 
\begin{figure}[htb]
\epsfig{figure=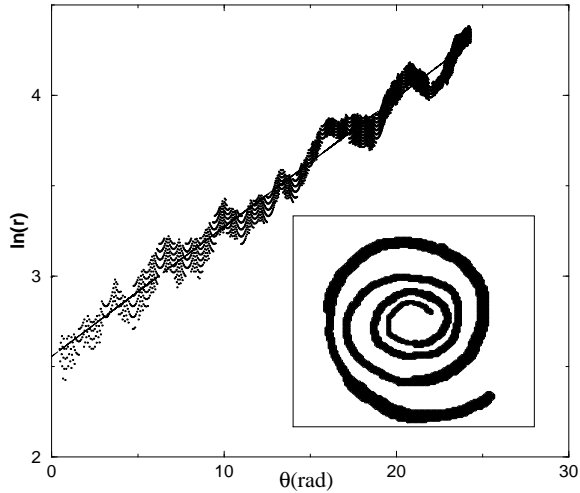,width=3.0in,angle=-0}
\caption{A spiral crack in ferric ferrocyanide precipitate and 
a fit of its trajectory in polar coordinates.
The linear fit has a slope of 0.072. 
}
\label{fig2}
\end{figure}
Read from the slope, the values of $k$
from 14 instances of spirals are summarized below,
for two different compounds and five different layer thickness.
For ferric ferrocyanide we got: $0.075$ (sample I.), $0.067$ (sample I.),
$0.080$ (sample I.), $0.082$ (sample I.), 
$0.086$ (sample I.), $0.059$ (sample II.),
$0.072$ (sample II.); for nickel phosphate: $0.078$ (sample III.), 
$0.075$ (sample III.), $0.063$ (sample III.), $0.061$ (sample III.), 
$0.061$ (sample IV.), $0.078$ (sample V.), $0.058$ (Sample V.)   
Some observations can be made at this point: 

(i) The logarithmic spiral, also known as the equiangular spirals,
has been studied extensively since the 17th century 
by Descartes, Torricelli and Jacques Bernoulli.
In addition to the well-known example in nautilus shells\cite{nautilus},
it also describes the shapes of the arms of some spiral galaxies 
such as NGC~6946 \cite{galaxy}, and 
the flight path of a peregrine falcon to its prey\cite{falcon}.

(ii) The apparent length-scale $r_0$ from equation (\ref{eq:r_theta}) 
can be absorbed in the phase:
$r(\theta)=e^{k(\theta-\theta_0)}$,
with $\theta_0=-ln(r_0)/k$. This demonstrates that the logarithmic spirals
are scale-free.

(iii) Surprisingly, the majority of experimentally measured
values of $k$ fall between 0.06 and 0.08. 
The fluctuation among different spirals of the same sample
is of the same magnitude as that
among different samples with different compounds and thickness. 
This indicates a mechanism 
independent of layer thickness and precipitate type 
for the formation of the observed spiral cracks. 

(iv) There are pronounced oscillations accompanying the linear trend 
in all our fittings, as exemplified in Fig.~2. 
These oscillations are almost periodical, and increases with $r$
(note the logarithmic scale.) 
\vspace{0.3in}
\begin{figure}[htp]
\epsfig{figure=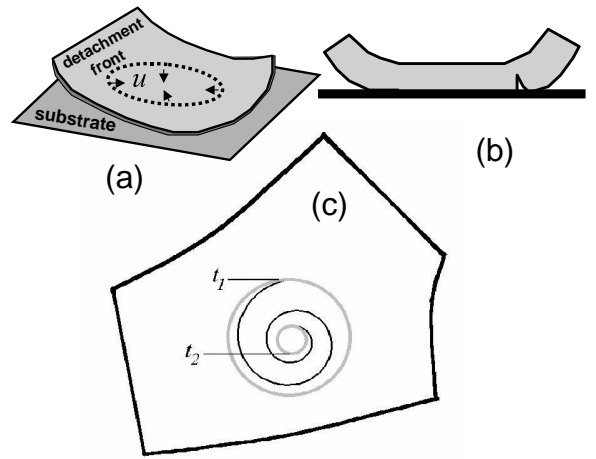,width=3.5in,angle=-0}
\caption{The proposed mechanism for generating spiral cracks:
(a) After primary fragmentation, a fragment folds up due to desiccation
with a shrinking detachment front,
(b) crack nucleation along the front and 
(c) the advancement of the front from time $t_1$ to $t_2$
leads to an inward propagating spiral crack.
}
\label{fig3}
\end{figure}
In-situ observations under microscope during the desiccation process
suggests the following mechanism. 
The spirals and circle-shaped 
structures are formed only after the primary fragmentation process
is over. Due to the humidity gradient across the thickness, the fragments 
gradually fold up and detach from the substrate (Fig.~3a-3b), 
generating large
tensile stress in the radial direction, at and normal to the front of
detachment. The extend of the attached area shrinks as the ring-shaped
front advances inward due to ongoing desiccation. When the stress at
the front exceeds the local material strength, 
a crack is nucleated (Fig.~3c).
Since the nucleation is seldom symmetrical with respect to the boundaries,
the crack tends to propagate along the front in one preferred
direction where more stresses can be released. 
In the absence of further nucleation event,
by the time the crack growth completes
a cycle the front has already advanced, 
forcing the crack to turn further inward, 
resulting eventually in a spiral crack (Fig.~3d). 
Since the stresses 
on the top of the layer are mostly relieved by folding, they are
concentrated at the layer-substrate interface. 
Therefore, the spiral runs
like a tunnel,
with $20-60\%$ penetration into the layer thickness. 
The patterns being largely spiral means that crack propagation is favored
over nucleation, for otherwise we would 
have observed more cylindrical concentric structures. 

We now investigate in more detail the conditions under which the 
logarithmic spiral cracks can be produced. Denote the speed of the
inward propagating drying front by $u$ and the speed of the crack tip by 
$v$. We choose the origin as the center of the spiral. The trajectory of
the crack is parameterized by the linear distance $s$, the radial distance
$r$, and the polar angle $\theta$, as shown in Fig.~4.

%\vspace{0.3in}
\begin{figure}[h]
\epsfig{figure=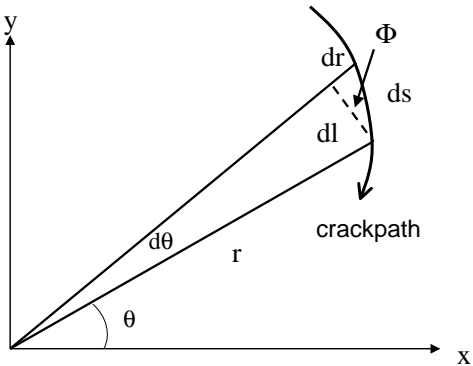,width=3.0in,angle=0}
\caption{Parametrization of the crack path.
}
\label{fig4}
\end{figure}

For infinitesimal $d\theta$, 
we have $dl=rd\theta$, $(ds)^2=(dr)^2+(dl)^2$, and $\tan \Phi=dr/dl$. Then:
\begin{eqnarray}
u^2 &=& \left( \frac{dr}{dt} \right)^2=\left( \frac{dr}{d\theta} \right)^2
\left ( \frac{d\theta}{dt} \right)^2, 
\label{eq:usq}
\\
v^2 &=& \left( \frac{ds}{dt} \right)^2=
        \left[ \left(\frac{dr}{d\theta}\right)^2 + r^2 \right]
            \left( \frac{d\theta}{dt} \right)^2.
\label{eq:vsq}
\end{eqnarray}
Eliminating $d\theta/dt$ from (\ref{eq:usq}) and (\ref{eq:vsq})
we obtain:
\begin{equation}
\frac{dr}{d\theta}=\frac{u r}{\sqrt{v^2-u^2}}=kr,
\label{eq:drdtheta}
\end{equation}
where
\begin{equation}
k\equiv \frac{u}{\sqrt{v^2-u^2}}=\tan\Phi.
\label{eq:k}
\end{equation}
For constant $k$, we get (\ref{eq:r_theta}), i.e., a logarithmic spiral.
Physically, since $k(u,v)$ is a function of $u/v$ alone,
this result implies that the general condition 
for a logarithmic spiral to occur is to have a constant 
ratio between the two velocities, despite the possibility that 
the actual dynamics may be very complicated and
neither speed is constant. 
In other words, the logarithmic spiral structure is 
surprisingly robust to dynamical details.
On the other hand since it is reasonable to expect a quasi-static crack 
to follow the direction of maximal stress relief,
the geometric result of a constant angle $\Phi$ 
suggests that the orientation of the maximal stress 
component is constantly bending away from the instantaneous
direction of propagation.
The fact that all fitted values of $k$ fall within a narrow range 
means that the degree of such bending is insensitive to details like the 
thickness or the chemical composition. 
More theoretically, one could ask if the Cotterell and Rice 
criterion \cite{CotterellRice} applies here: whether 
the stress intensity factor, $K_{II}$,
vanishes along a constant inclination as the crack propagates 
under the influence of a radially symmetric stress field. 
If so, a logarithmic spiral is naturally expected. 

Now we turn our attention to the oscillations that decorate
the logarithmic behavior, as displayed in Fig.~2. 
Such oscillations indicate the departure from radial symmetry in the 
underlying stress field.
It can be explained on the basis of boundary effects.
Clearly the shape of a typical fragment is generally
polygonal, not circular, and so its free boundaries 
will modify the shape of the shrinking stress front,
more so the closer the front is to the boundaries.
This leads to periodic variations in $\Phi$, and hence oscillation in $k$. 
For rectangular fragments we expect to see two cycles 
of oscillation per revolution of the spiral. 
This is indeed roughly the case in our fits. 
Moreover, the observed diminishing amplitude of oscillation 
at smaller $r$ is also in agreement. 

To confirm and test the robustness of our proposed 
mechanism, we implement it in a mesoscopic spring-block model \cite{prl}
which describes fracture of an overlayer on a frictional substrate.
In this model, the grains in the layer are
represented by blocks which form a triangular array of linear size $L$
with interconnecting bundles each of which has $H$ bonds 
(Hookean springs).
The bond has a breaking strength $F_c$ and a relaxed length $l$. 
While $H$ plays the role of thickness, the initial block spacing 
$a$ prescribes the residual strain $s=(a-l)/a$. Each block has the
same slipping threshold $F_s$ such that whenever the magnitude of the 
resultant force $\vec{F}$ on a block exceeds $F_s$, the block 
slips to an equilibrium position defined by $\vec{F}=0$. 
The temporally increasing stress in the layer during desiccation 
can be modeled by increasing the stiffness of the bonds. 
This is however equivalent to the case of constant stiffness 
but decreasing $F_s$ and $F_c$, with fixed $\kappa\equiv F_c/F_s$.
In this way, the competition between 
stick-and-slip and bond breaking, quantified by the
set of parameters $\Gamma=\{s,\kappa,H,L\}$, 
was shown to give rise to realistic fragmentation 
and select the emerged scale\cite{prl}.

Here we focus instead on what happens after the fragmentation process
has settled. The simulated system is thus assumed to represent a 
fragment stable against the primary fragmentation 
(ensured by choosing $\Gamma$ properly), but susceptible
to secondary curved crackings induced by an advancing stress front.
Therefore, an inhomogeneous stress field is imposed,
with a profile constant everywhere 
at $\sigma_0$ except on an annular region of radius $R$,
where there is a hump of height $\Delta \sigma$ and width $w$
(see upper inset of Fig.~5). 
By decreasing $R$ at a constant speed $u$, we model the advancing front
of stress field caused by detachment. In a rather narrow parameter region
(mainly small $u$, large $\kappa$, $\Delta \sigma\sim\sigma_0$ 
and $w\approx a$),
the desired spiral cracks are successfully reproduced 
(lower inset of Fig.~5).
Consistent with experiment, the simulated spirals 
also follow a logarithmic form as illustrated in Fig.~5. 
The value of $k$ depends on the parameters of the model. 
In particular, imposing smaller penetration results in
tighter-binding spirals and hence smaller $k$. 
This is consistent with the screening effects of 
the existing crack on the stress field that influences
further propagation of the crack.

\begin{figure}[htb]
\epsfig{figure=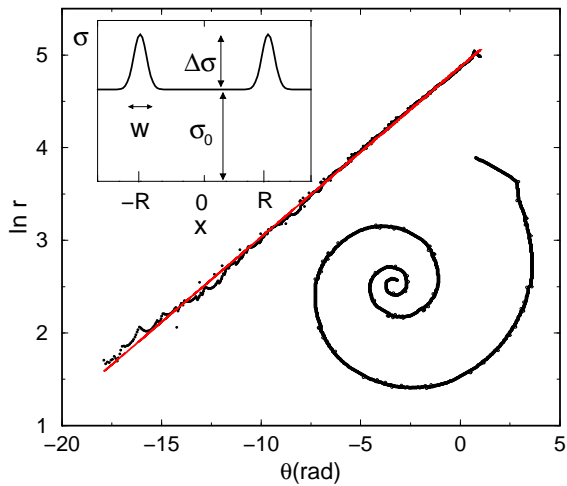,width=3.5in,angle=-0}
\caption{
A simulated spiral crack (lower inset) and a fit of its trajectory.
The simulation parameters are:  $L=300$, $s=0.3$, $\kappa=2$, $H=4$, 
$\Delta \sigma/\sigma_0=1.4$, $w=a$, speed $u=0.005a$/step, 
and full penetration.
The linear behavior indicates a logarithmic spiral,
with a slope$\approx 0.184$.
Upper inset: schematic plot of the stress profile imposed on the fragment.
}
\label{fig5}
\end{figure}

In conclusion, spiral crack, an astonishing member of the family
of fascinating patterns produced by fracture, 
is realized in a surprisingly simple setup of desiccating precipitate.
Their formation is argued to be driven by an unusual stress
relaxation process governed by the fold-up of the fragments. 
A discrete spring-block model incorporating such a driving
appears to capture successfully the observed phenomena. 
The spirals have interesting properties,
of which several quantitative aspects can be understood theoretically.
 
The work of K.-t.Leung is supported by the National Science Council of R.O.C. and 
the work of Z. Neda is sponsored by the Bergen Computational Physics Laboratory.
We are grateful for the professional advices and comments from 
Professor Derek Hull.

%\vspace{1.5cm}
%:::::::::::::::::::::::::::::::::::::::::::::::::::::::::::::::::::::::
%References:
%:::::::::::::::::::::::::::::::::::::::::::::::::::::::::::::::::::::::


\begin{references}

\bibitem{fra_gen} B. Lawn, Fracture of brittle solids, 2nd ed. (Cambridge
Univ. Press, New York, 1993); D. Hull, Fractography (Cambridge Univ. Press,
Cambridge, 1999) 

\bibitem{recent} A. Yuse and M. Sano, Nature {\bf 362}, 329 (1993);
A. Groisman an E. Kaplan, Europhys. Lett. {\bf 25}, 415 (1994); 
T. Bai, D.D. Pollard and H. Gao, Nature {\bf 403}, 753 (2000);
K.A. Shorlin, J.R. de Bruyn, M. Graham and S.W. Morris, Phys. Rev. E  {\bf 61},
6950 (2000).

\bibitem{models} B. K. Chakrabati and L.G. Benguigui, Statistical Physics
of Fracture and Breakdown in Disordered Systems (Clarendon Press, Oxford, 1997);
A.T. Skjeltorp and P. Meakin, Nature {\bf 335}, 424 (1988); 
K.-t. Leung and J.V. Andersen, Europhys. Lett. {\bf 38}, 589 (1997).

\bibitem{curving} J. K. Gillham,P. N. Reitz and M.J Doyle, Polymer Eng. Sci.
{\bf 8}, 227 (1968).

\bibitem{tibia} O.M. Bostman, J. Bone Joint Surg. {\bf 68}, 462 (1986).

\bibitem{pipes} 
Natural Gas Pipeline Rupture, Pipeline Occurrence Report Number P96H0012,
The Transportation Safety Board of Canada,
http:// www.tsb.gc.ca/ ENG/ reports/ pipe/ 1996/ p96h0012/ ep96h0012.html.

\bibitem{Hull} D. Hull, Fractography, Chapter 3: Tilting cracks pp.70 
(Cambridge Univ. Press, Cambridge, 1999).

\bibitem{nature}
K.-t. Leung, L. J{\'o}zsa, M. Ravasz and Z. N{\'e}da,
%{\it Spiral cracks without twisting\/},
Nature {\bf 410}, 166 (2001).

\bibitem{xia}
Z.C. Xia and J.W. Hutchinson, 
J. Mech. Phys. Solids, {\bf 48}, 1107 (2000)

\bibitem{nautilus}  
D'Arcy Thompson, On Growth and Form (Cambridge Univ. Press, Cambridge, 1961).

\bibitem{galaxy}
P. Frick et al., Mon. Not. R. Astron. Soc. {\bf 318}, 925 (2000)

\bibitem{falcon} 
A. E. Tucker, K. Akers and J. H. Enderson,
J. Exp. Biol. {\bf 203}, 3733; 3745; 3755 (2000).

\bibitem{CotterellRice}
B. Cotterell and J.R. Rice, Int. J. Fract. Mech. {\bf 16}, 155 (1980).

\bibitem{prl}
K.-t. Leung and Z. N\'eda, Phys. Rev. Lett. {\bf 85}, 662 (2000).

\end{references}
\end{document}